\documentclass[aps,prl,showpacs,showkeys,reprint,groupedaddress,notitlepage]{revtex4-1}
\pdfoutput=1
\usepackage{siunitx}
\usepackage{amsmath}
\usepackage{graphicx}

\newcommand{\elok}{\widetilde{E}_{LO}}
\newcommand{\esk}{\widetilde{E}_S}
\newcommand{\elo}{E_{LO}}
\newcommand{\es}{E_S}

\begin{document}

\title{Multimode Quantum State Tomography Using Unbalanced Array Detection}
\author{Brandon S. Harms}
\author{Blake E. Anthony}
\author{Noah T. Holte}
\author{Hunter A. Dassonville}
\author{Andrew M.C. Dawes}
\email[]{dawes@pacificu.edu}
\homepage[]{www.amcdawes.com}
\affiliation{Pacific University, Department of Physics, Forest Grove Oregon USA 97116}

\date{\today}

\begin{abstract}
We measure the joint Q-function of a multi-spatial-mode field using a
charge-coupled device array detector in an unbalanced heterodyne configuration.
The intensity pattern formed by interference between a weak signal field and a
strong local oscillator is resolved using Fourier analysis and used to
reconstruct quadrature amplitude statistics for 22 spatial modes simultaneously.
The local oscillator and signal propagate at an angle of \SI{12}{\milli\radian}
thus shifting the classical noise to modes that do not overlap with the signal.
In this configuration, balanced detection is not necessary.
\end{abstract}

\pacs{42.50.Ar, 03.65.Ta, 03.65.Wj}
\keywords{tomography, Q-function}

\maketitle

Multimode quantum optical systems are increasingly important in applications
such as slow and stored light devices \cite{Grodecka-Grad:2012aa}. A complete
understanding of these devices requires a fidelity measurement of the stored
quantum state, but there has not yet been a simple way to measure the quantum
state of a multi-spatial mode field. Balanced homodyne detection (BHD) is the
standard measurement technique used to construct the complete quantum mechanical
state of light \cite{leonhardt_measuring_1997}. This technique has one primary
weakness: multimode fields or fields in an unknown spatial mode suffer
significant losses due to mode mismatch between the local oscillator (LO) and
the signal field \cite{leonhardt_measuring_1997,beck_quantum_2000}.

Balanced array detection overcomes mode-matching losses and has been used to
perform measurements of more than two modes at once
\cite{raymer_many-port_1993}. It is possible to obtain simultaneous (but not
joint) measurements of the Wigner functions of many modes by using array
detectors \cite{dawes_simultaneous_2001,dawes_mode_2003}. It has also been shown
theoretically that array detectors may be used to measure the joint Q-function
of a multimode field \cite{iaconis_temporal_2000}. One of the difficulties of
these methods is the need to balance two array detectors and perform
pixel-by-pixel subtraction to eliminate the classical intensity noise of the LO
\cite{Yuen:1983aa}. These alignment requirements become prohibitive when trying
to measure two transverse directions. This difficulty was removed by Beck et
al.\ who achieved the same effect with a single array, using unbalanced
detection, by arranging the LO and signal fields to isolate the classical LO
noise \cite{Beck:2001aa}.

In this Letter, we describe a multi-spatial-mode detection method that is
analogous to the multi-temporal-mode detection method of Beck et al. We
simultaneously measure many spatial modes without having to vary the propagation
direction or the phase of the LO; this reduces the amount of data that needs to
be acquired and enables simultaneous joint measurements of two modes. This
measurement technique can measure joint statistics between two independent
spatial modes. Knowledge of these inter-mode correlations could provide a new
route to information storage in a wide variety of systems.

To develop the theory of unbalanced array detection, we consider the spatial
intensity $S(\mathbf{x})$ incident on the array detector due to the
combination of an LO field $\elo$ and the signal field $\es$. For simplicity,
we describe only one transverse coordinate (i.e. horizontal); generalization
to both transverse coordinates is straightforward. If the LO is normally
incident on the detector and the signal field propagates at a small angle to
the LO, the spatial intensity at the detector is given by
\begin{align}
  S(\mathbf{x})&=|\elo(\mathbf{x})+\es(\mathbf{x})\exp(i\mathbf{k_S\cdot x})|^2\nonumber\\
  &=|\elo(\mathbf{x})|^2+|\es(\mathbf{x})|^2\nonumber\\
  &+\left[{E}^*_{LO}(\mathbf{x})\es(\mathbf{x})\exp(i\mathbf{k_S\cdot x})+ c.c.\right].
\end{align}

The Fourier transform of this measurement yields
\begin{align}
  \widetilde{S}(\mathbf{k})&=\elok^*(\mathbf{-k})\otimes\elok(\mathbf{k})+
\esk^*(\mathbf{-k})\otimes\esk(\mathbf{k})\nonumber\\
  &\quad + f(\mathbf{k-k_S})+f^*(\mathbf{-k-k_S}),
\end{align}
where $f(\mathbf{k}) = \elok^*(\mathbf{-k})\otimes\esk(\mathbf{k})$ and
$\otimes$ denotes the convolution. The first of the three terms peaks at
$\mathbf{k=0}$ and contains the second-order classical LO noise that would be
removed in balanced detection. The second term is negligible if the signal field
is weak. If $\mathbf{k_S}$ is large enough, the function $f$ contains all
information on the heterodyned signal. This function has a peak value at
$\mathbf{k=k_S}$ and is therefore separate from the classical noise at
$\mathbf{k}=0$. Just as Beck et al.~\cite{Beck:2001aa} eliminate classical LO
noise by temporally separating the signal and LO fields, we eliminate classical
LO noise by separating the LO and signal fields by their propagation
direction---they have different transverse components of their propagation wave
vector \cite{Dawes:2013aa}.

\begin{figure}[htbp]
    \includegraphics[scale=0.75]{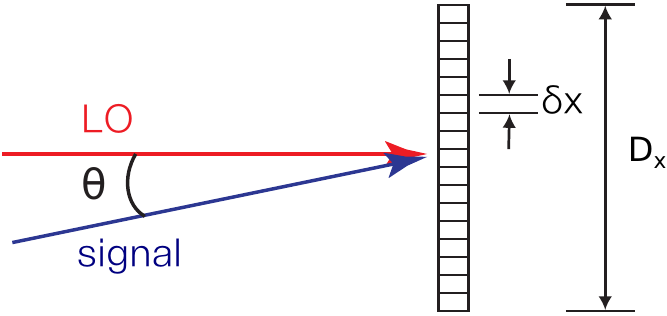}
    \caption{A normally-incident plane-wave local oscillator (LO) interferes
    with a signal field at angle $\theta$. The array detector pixels measure $\delta x$ and
    the detector size is $D_x$.}
    \label{fig:losig}
\end{figure}

To describe the detection process from the perspective of quantum mechanics, we
consider the spatial pattern detected at an array, as  in Fig.~\ref{fig:losig},
where $\hat{n}_j=\hat{a}^\dagger_j\hat{a}_j$ is the  operator corresponding to
the number of photons incident on pixel $j$ of the  array. Each pixel measures a
different spatial position $x_j=j\delta x$ and  there are $N$ pixels labeled by
$-N/2 \leq j < N/2$. We can also express the  measured field in terms of
plane-wave modes $\hat{b}_k$, $-N/2\leq k<N/2$ using  the Fourier relation
\begin{align}
  \label{a_to_b}
  \hat{a}_j = \sum_k \exp\left[-i2\pi j k/N\right]\hat{b}_k
\end{align}
We assume the LO occupies the $2M+1$ plane-wave modes near the center of the
spectral window. The signal occupies the plane-wave modes with positive $k$ and
the plane-wave modes with negative $k$ contain only the vacuum field. This is
explicitly written as
\begin{equation}
  \label{3bs}
  \hat{b}_k = \begin{cases}
    \hat{b}_k^{(vac)} & -N/2\leq k < -M,\\
    \hat{b}_k^{(LO)} & -M\leq k \leq M,\\
    \hat{b}_k^{(s)} & M < k < N/2.
  \end{cases}
\end{equation}
The operator we measure corresponds to the inverse Fourier transform of
$\hat{n}_j$
\begin{align}
  \label{K_to_n}
  \hat{K}_p = \frac{1}{\sqrt{N}} \sum_j \exp\left[i2\pi p j/N\right]\hat{n}_j
\end{align}
Combining Eqns.~\ref{a_to_b} and \ref{3bs} yields an expression for the
per-pixel number operator \(\hat{n}_j\) in terms of plane-wave mode operators
\(\hat{b}_k\). Terms that combine two weak fields \(\hat{b}_k^{(vac)}\) and
\(\hat{b}_k^{(s)}\) are discarded as second-order. From here, Eqn.~\ref{K_to_n}
is used to relate \(\hat{K}_p\) in terms of the \(\hat{b}_k\)'s. Finally, the
large-amplitude of the LO field means the LO operators \(\hat{b}_k^{(LO)}\) can
be replaced with their coherent state amplitudes $\beta_k$. For $p > 2M$ we find
\begin{equation}
  \label{K_to_b}
  \hat{K}_p = \sum_{k=-M}^{M}\left( \beta_k^*\hat{b}^{(s)}_{k+p}+
  \beta_k\hat{b}^{\dagger(vac)}_{k-p}\right).
\end{equation}

If the LO occupies only a single ($k=0$) plane-wave mode, then Eq.~\ref{K_to_b}
simplifies to
\begin{equation}
  \label{eqn:K_to_b2}
  \hat{K}_p = \beta_0^*\hat{b}^{(s)}_{p}+ \beta_0\hat{b}^{\dagger(vac)}_{-p}.
\end{equation}
For an LO in a single plane-wave mode, a measurement of $\hat{K}_p$ (the Fourier
transform of the photocount data) returns a complex number. Using
Eq.~\ref{eqn:K_to_b2}, this complex number can be interpreted as a measurement
of $\hat{b}_p^{(s)}$, the signal mode annihilation operator, plus a vacuum
contribution (the second term in Eq.~\ref{eqn:K_to_b2}). The annihilation
operator is itself the sum of the two field quadrature amplitudes
\begin{equation}
    \label{eq:quadratures}
  \hat{b}_p^{(s)} = \frac{1}{\sqrt{2}}\left(\hat{x}_p+i\hat{y}_p\right).
\end{equation}
Therefore, the real and imaginary components of each $\hat{K}_p$ correspond to
simultaneous measurement of the quadrature amplitudes $x_p$ and $y_p$. Of course
these observables are noncommuting so the ability to measure them simultaneously
comes at the price of additional vacuum noise
\cite{arthurs_simultaneous_1965,leonhardt_measuring_1997}. It is precisely this
additional noise that prevents reconstruction of the Wigner function, instead
allowing reconstruction of the Q-function.

It is important to note that each Fourier-transformed exposure of the array
returns a set of $(N/2) - M$ complex numbers. Each of these, indexed by $p$, has
a real and imaginary part that corresponds to the signal mode field quadrature
$(x_p,y_p)$. If one value of $p$ is selected, and the corresponding $(x_p,y_p)$
pairs are histogrammed, the result tends toward the Q distribution for the field
quadratures of mode $p$. Because the data for all values of $p$ are collected
simultaneously, any mode $p$ can be analyzed from a single set of data.
Additionally, joint Q-distributions can be computed for any pair or combination
of modes.

\begin{figure}[htbp]
    \includegraphics[width=8cm]{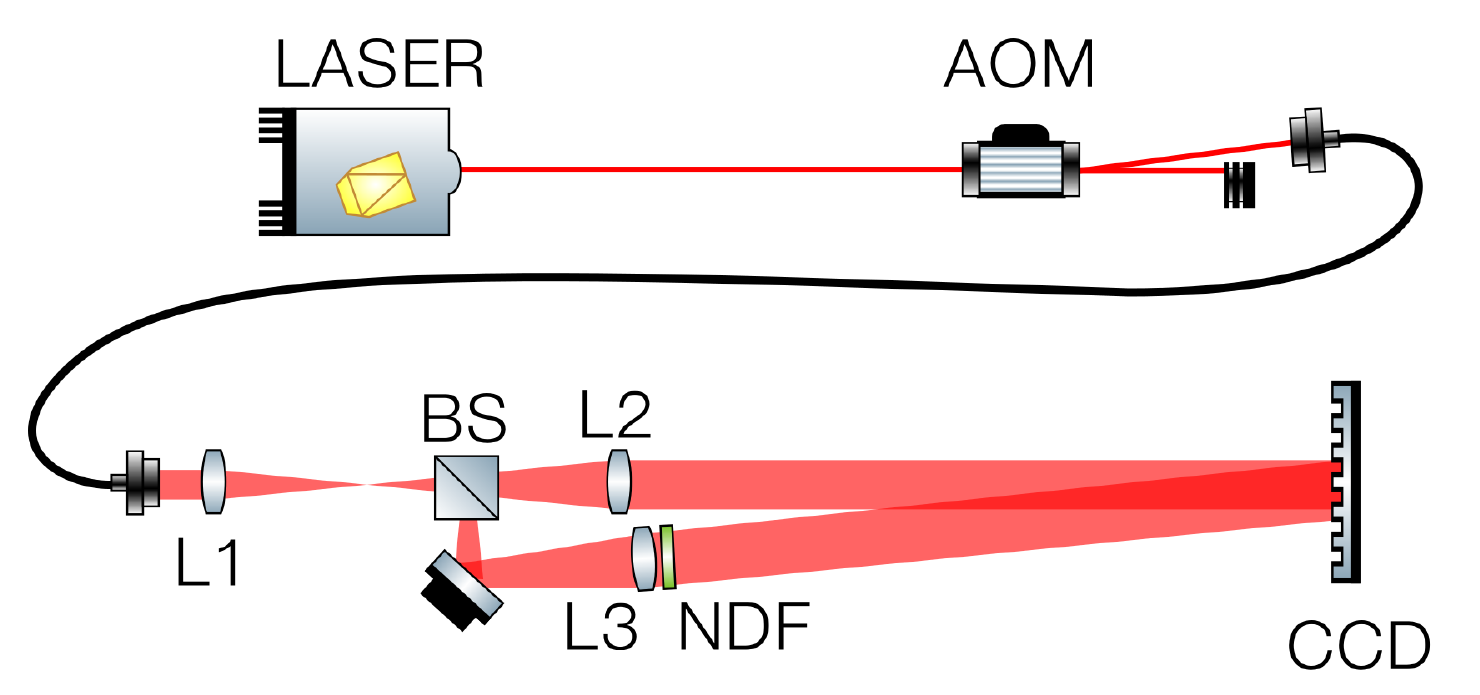}
    \caption{Experimental setup: a 780 nm external cavity diode laser is coupled
    into single-mode fiber, gated using an 80-MHz acousto-optic modulator (AOM),
    and split into LO and signal fields that interfere on a CCD array. Lens
    focal lengths are: L1 50 mm, L2 250 mm, L3 300 mm.}
    \label{fig:setup}
\end{figure}

Our implementation of this system, shown in Fig.~\ref{fig:setup}, begins with a
\SI{780}{\nano\meter} external-cavity diode laser. The laser output is sent
through an acousto-optic modulator (80 MHz). The first order diffracted beam is
then spatially filtered using an optical fiber (780-HP). After the fiber output,
a telescope, starting with a 50-mm focal length lens, expands the beam diameter
to \SI{1}{\centi\meter}. A 50/50 beam splitter separates the LO from the signal
and a 250-mm focal length lens collimates the LO beam after expansion. The
signal beam remains slightly diverging after passing through a 300-mm focal
length lens. The mode mismatch between the signal and the LO provides a simple
multimode signal for our demonstration of this method. The signal is attenuated
by a factor of $10^5$ using neutral density filters. The signal and LO interfere
on a charge coupled device (CCD) camera \footnote{Princeton Instruments, PyLoN
400BR eXcelon.}. The CCD is a 1340$\times$400 array of $\delta x
=$~\SI{20}{\micro\meter} square pixels with quantum efficiency of 98\% at
\SI{780}{\nano\meter}, cooled to \SI{-120}{\celsius} to achieve a dark current
rate of 2-3 electrons per pixel per hour.

The LO power incident on the CCD is $\sim$\SI{1}{\micro\watt} and the
LO and signal beams interfere on the CCD at an angle of
$\sim$\SI{12}{\milli\radian}. The AOM signal is modulated by a function
generator (Rigol DG4062) to create a \SI{7.5}{\milli\second} rectangular pulse.
The pulse is triggered \SI{5}{\milli\second} after the start of the
\SI{20}{\milli\second} exposure to keep the CCD dark during readout.

The largest noise source in our measurements is the electronic readout noise. We
characterize this noise by measuring the variance in photocounts $\Delta n_j$
with the LO illuminating the CCD and without illumination. The variance with
illumination is an average of \SI{15}{\deci\bel} above the variance without
illumination so our signal to noise ratio is more than adequate.

If the signal is blocked, the signal mode entering the detector is then the
vacuum mode, and the quadrature amplitudes measured should be zero. Despite our
spatial filtering, the residual spatial components in the LO beam leave a
background signal in the FFT output that should be subtracted before computing
the Q-function. To carry out this correction, we collect 500 exposures with the
signal blocked (i.e. the signal in the vacuum state) and average the FFT output
across these 500 exposures. This vacuum average $\langle
\hat{K}_{p}\rangle_\mathrm{vac}$ is then subtracted from subsequent data prior
to further calculations.

We collect 1000-8000 shots of CCD data from an ROI at the center of the CCD (600
pixels wide by 10 pixels tall), sum vertically the ten rows to obtain $N =
600$ data points and compute the FFT of the resulting 600-element array. The
output of this calculation is an array of 600 complex values of $K_p$ where the
first 300 values correspond to unique modes indexed by $p$ \footnote{Given that
the raw CCD data is real, the FFT output is symmetric about the midpoint.}.

\begin{figure}[htbp]
    \includegraphics[width=8.6cm]{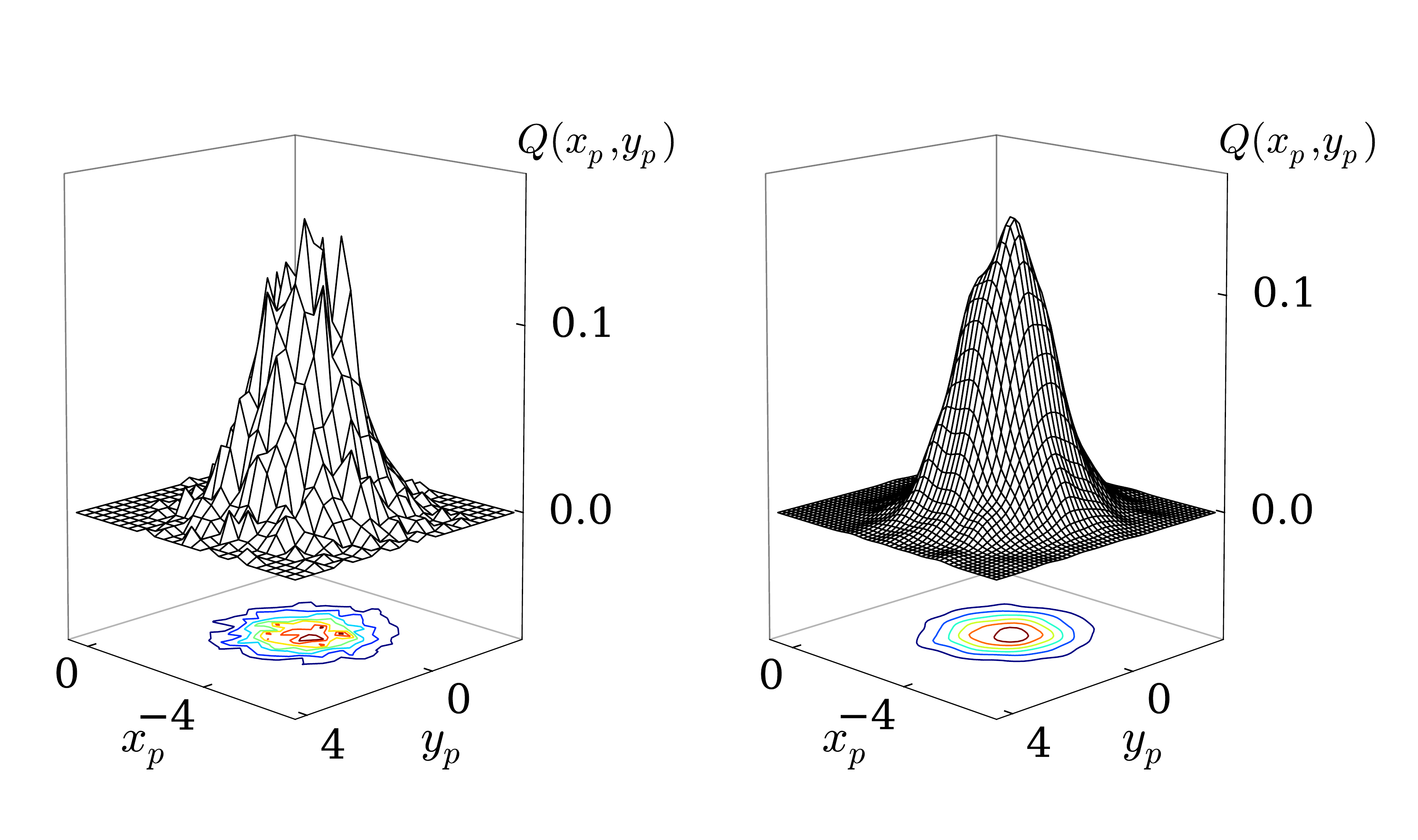}
    \caption{Figure of single mode Q-function. The mode selected here has an
    average of $\langle n \rangle =
7.2$ photons and $\Delta n=2.8$, within 4\% of the quantum noise limit. Shown
on the left is the raw-data histogram and on the right is a
kernal density estimate calculated from the raw data.}
    \label{fig:qfunc}
\end{figure}

We compute the quadratures by scaling the data by the total number of photons
$n_t$ detected per shot in the CCD region of interest. For weak signal fields,
$n_t$ is essentially equal to the average number of photons in the LO and
therefore related to the classical LO amplitude as $$|\beta| = \sqrt{n_t}.$$
Therefore, the quadratures in Eq.~\ref{eq:quadratures} are calculated from
experimental values of $K_p$ as
\begin{align}
    \label{eq:quads}
    x_p = \sqrt{\frac{2}{n_t}}\mathrm{Re}\left(K_p - \langle K_p \rangle_{vac}\right),\\
    y_p = \sqrt{\frac{2}{n_t}}\mathrm{Im}\left(K_p - \langle K_p \rangle_{vac}\right).\\
\end{align}

From the quadratures, we compute the Q-function \(Q(x_p,y_p)\) for a specific
mode $p$. This calculation is done in two ways. First, we simply histogram the
$x_p$ and $y_p$ pairs using 30 bins in each dimension. For large data sets, this
histogram will approximate \(Q(x_p,y_p)\). Additionally, we calculate
\(Q(x_p,y_p)\) by applying a kernel density estimate (KDE) to the $x_p$ and
$y_p$ pairs. This method essentially places a narrow Gaussian function at each
point $(x_p,y_p)$ and computes the sum. The width of the Gaussian kernel is
determined using Scott's rule and results in some smoothing relative to our raw
histogram \cite{Scott:2009aa}.

\begin{figure}[htbp]
    \includegraphics[width=8.6cm]{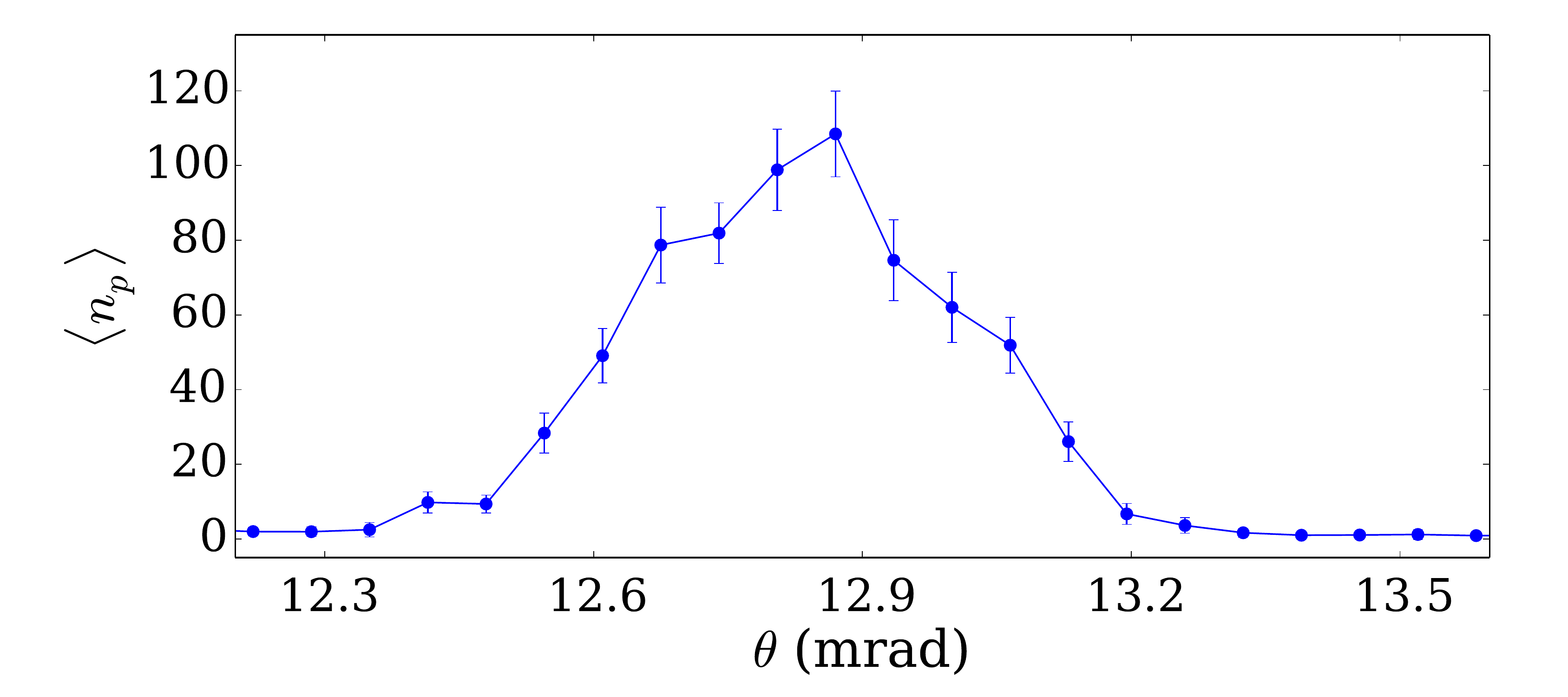}
    \caption{Average photon number vs. plane wave mode angle. For the 15 modes occupied by the
signal field we plot $\langle n_p \rangle$ with error bars given by $\Delta
n_p$.}
    \label{fig:nvp}
\end{figure}

In the following figures, both the raw-data histogram and the KDE versions of
the Q-function are presented for comparison. Shown in Fig.~\ref{fig:qfunc} is
$Q(x_p,y_p)$ for a mode with $\langle n \rangle = 7.2$ photons and $\Delta n =
2.8$. The Q-function can be computed for any mode $p$ using the same set of
data. Because data is collected for all modes simultaneously, we can also
compute a quantity of interest (such as average photon number $\langle n_p
\rangle$ and measure that quantity for all modes simultaneously. In
Fig.~\ref{fig:nvp} we plot average photon number $\langle n_p \rangle$ vs. plane
wave mode angle $\theta_p = p \lambda / (N \delta x)$. This illustrates the mode
spectrum of the signal field. In particular, with a multimode field there are
several modes with comparable amplitudes. While Fig.~\ref{fig:nvp} shows the 22
modes near the signal, we measure 600 modes simultaneously; highly
multimode signals can be measured using this technique.

\begin{figure}[htbp]
    \includegraphics[width=8.6cm]{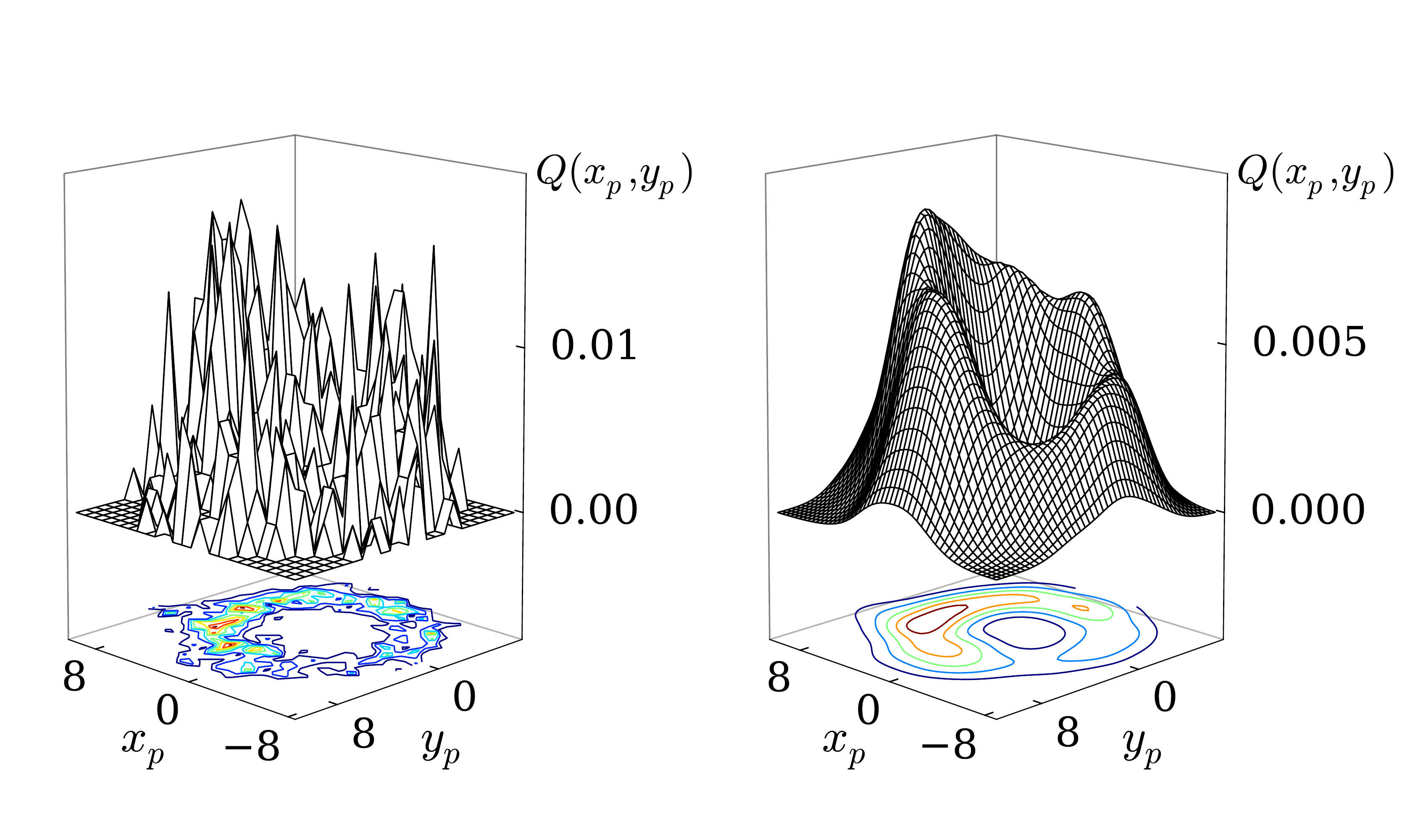}
    \caption{Q-function for mode with randomized phase. With randomized phase, the
Q-function takes on a donut shape centered at the origin. The consistent radius
indicates a stable photon number while the angular spread indicates the changing
phase.}
    \label{fig:randphase}
\end{figure}

To demonstrate the detection of a signal with variable phase, we dither the
position of a piezo-mounted mirror in the signal beam path and observe
modulation through \SI{2\pi}{\radian} of phase. The Q-function calculated from
such data is shown in Fig.~\ref{fig:randphase}. The characteristic donut shape
corresponds to phase noise although the phase modulation is not completely
random as evidenced by the peaks present in the Q-function.

\begin{figure}[htbp]
    \includegraphics[width=8.6cm]{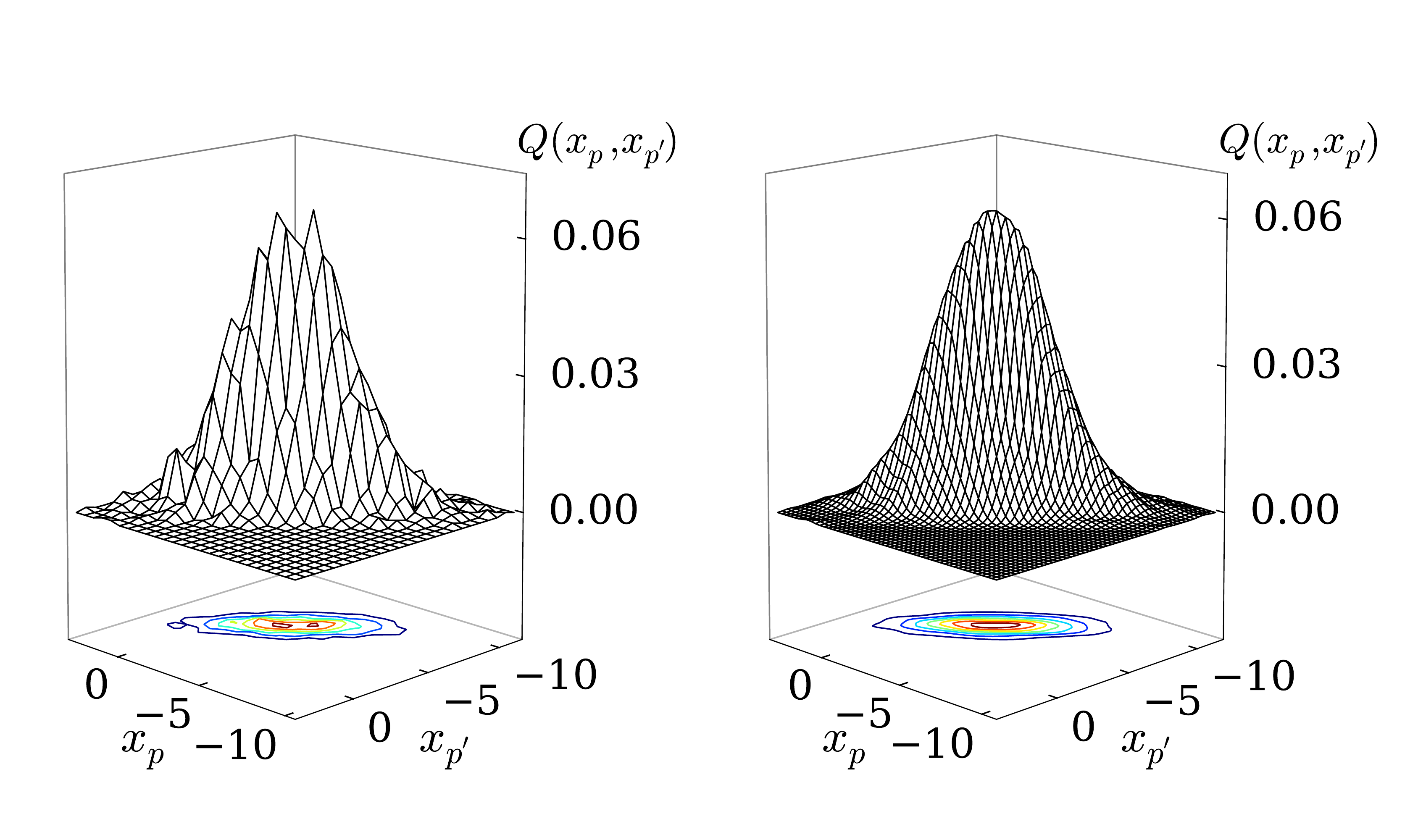}
    \caption{Two-mode joint Q-function. For adjacent modes ($p=197$ and $p'=198$) the
joint Q-function illustrates the correlation between the real parts, $x_{p}$, of
each mode.}
    \label{fig:joint}
\end{figure}

It is important to note that we can also compute any quantity that is a function
of the quadratures $x_p$ and $y_p$. One such quantity is the joint Q-function,
$Q(x_p,x_{p'})$. The joint Q-function shown in Fig.~\ref{fig:joint} is for two
nearby modes $p$ and $p'=p+1$. As expected, these modes exhibit strong positive
correlation between $x_p$ and $x_{p'}$.

In conclusion, we point  out that this method could be used to measure the
quantum state of light stored in a multimode quantum memory such as that
described in \cite{Grodecka-Grad:2012aa}. The peak sensitivity of our CCD system
is near the \SI{780}{\nano\meter} and \SI{795}{\nano\meter} resonances of
Rubidium. Very recently, quantum state tomography measurements have been
performed on single-photon states retrieved from a stored-light system
\cite{Bimbard:2014aa}. Multimode measurements of such systems could be made
using methods presented in this Letter and would yield new information about
correlations between modes retrieved from stored light systems.

\acknowledgments

We thank M. Beck for helpful discussions. This material is based upon work
supported by the National Science Foundation under Grant No. 1205828. Additional
financial support was provided by the Research Corporation for Science
Advancement and the Pacific Research Institute for Science and Mathematics.

\bibliography{unbalQST}

\end{document}